\documentclass[doublecol]{epl2} 

\usepackage{graphicx}

\title{Does the $8-N$ bonding rule break down in As$_2$Se$_3$ glass?}
\shorttitle{Does the $8-N$ bonding rule break down in As$_2$Se$_3$ glass?} %Insert here a short version of the title if it exceeds 70 characters

\author{S. Hosokawa\inst{1,2} \and A. Koura\inst{1} \and J.-F. B\'{e}rar\inst{3} \and W.-C. Pilgrim\inst{2} \and S. Kohara\inst{4} \and F. Shimojo\inst{1}}
\shortauthor{S. Hosokawa \etal}

\institute{                    
  \inst{1} Department of Physics, Graduate School of Science and Technology, Kumamoto University, Kumamoto, 860-8555, Japan\\
  \inst{2} Faculty of Chemistry, Physical Chemistry, Philipps University of Marburg, 35032 Marburg, Germany\\
  \inst{3} Institut N\'{e}el, Centre National de la Recherche Scientifique/Universit\'{e} Joseph Fourier (CNRS/UJF), 38042 Grenoble Cedex 9, France\\
  \inst{4} Research and Utilization Division, Japan Synchrotron Radiation Research Institute (JASRI/SPring-8), Hyogo 679-5198, Japan
}

\abstract{
The local coordination numbers of As$_2$Se$_3$ glass were determined by a combination of anomalous x-ray scattering experiments, reverse Monte Carlo calculations, and {\it ab initio} molecular dynamics simulations. The well-known `8-$N$ bonding rule' proposed by Mott breaks down around the As atoms, exceeding the rule by 7--26\%.  An experimental prediction based on mean-field theory agrees with the present experimental and theoretical results. The fourfold coordinated As atoms likely form As-As wrong bond chains rather than ethan-like configurations, which is identified as the origin for the breakdown of the  `8-$N$ bonding rule'.
}

\begin{document}

\maketitle

%\section{Introduction}
It was widely believed that coordination numbers of covalent bonding glasses were governed by the `8-$N$ bonding rule', where $N$ is the total number of $s$ and $p$ electrons in the outmost shell of atoms. This rule was proposed by a Novel prize winner, N. F. Mott \cite{Mott1969}, who emphasized that according to this rule, all electrons are situated in filled electronic bands so that large changes of the electrical conductivity do not occur in covalent glasses with a small change of composition \cite{MottBook}, in contrast to crystalline semiconductors. 

Street and Mott \cite{StreetMott} proposed a model of dangling bond pairs in amorphous chalcogenide semiconductors to explain the pinning effect of the Fermi level. Subsequently, Kaster et al. \cite{Kastner} proposed another model of valence alternative pairs where threefold and dangling bond Se atoms appear as defect centers. These models showed a breakdown of the `8-$N$ bonding rule'. However, these defects discussed there are in a negligible concentration regime undetectable by scattering experiments although they are very important for considering electronic states of glasses.

To our knowledge, only one group expressed their opposition to this rule experimentally \cite{Georgiev}. Their work was carried out on As$_x$Se$_{1-x}$ glasses using temperature-modulated differential calorimetry. It was found that non-reserving heat flow almost vanishes in the $0.29<x<0.37$ composition range, separating the floppy glasses from the stressed rigid ones within the framework of the rigidity percolation theory \cite{Phillips, Thorpe}. Since the ideal boundary should be $x=0.40$ if the 8-$N$ bonding rule was strictly valid, they proposed a breakdown of this rule and suggested that 28.6\% of Se=As(Se$_{1/2}$)$_3$ quasi-tetrahedral units should exist in addition to the normal As(Se$_{1/2}$)$_3$ pyramidal units around the As atoms. This idea motivated diffraction and spectroscopic investigations of the local structures in these glasses. However, no experimental evidence could be achieved over the last ten years due to the experimental difficulties in determining local structures of glassy systems.

In this letter, we present an experimental evidence for the breakdown of the `8-$N$ bonding rule' around the As atoms in the As$_2$Se$_3$ glass obtained from a combination of anomalous x-ray scattering (AXS) experiments and reverse Monte Carlo (RMC) calculations. This result is theoretically supported by an {\it ab initio} molecular dynamics (MD) simulation.

%\section{Experimental procedure}
The bulk glass As$_2$Se$_3$ sample was obtained by quenching the melt in a quartz ampoule containing the mixed compound. The melt was homogenized at 600$^\circ$C for at least 48 h before the sample was slowly quenched in air. 

The AXS experiments were performed at two energies below each $K$ edge (20 and 200 eV below the As $K$ edge of 11868 eV and the Se $K$ edge of 12658 eV) at the beamline BM02 of the European Synchrotron Radiation Facility (ESRF) in Grenoble, France. The diffraction measurements were carried out in reflectance geometry using a standard $\omega-2\theta$ diffractometer installed at the beamline. For discriminating the elastic signal from the $K\beta$ fluorescence and Compton scattering contributions, a graphite crystal analyzer was mounted on a 1-m-long detector arm. The feasibility of this detecting system was described in detail elsewhere \cite{HosokawaAIP, HosokawaGeSe}. Following the procedure given in Ref. \cite{HosokawaZPC}, differential structure factors, $\Delta_kS(Q)$, were obtained from the difference between diffraction data with two different incident x-ray energies near each $K$ edge of $k$ element.

The RMC simulations were performed using $\Delta_{\rm As}S(Q)$, $\Delta_{\rm Se}S(Q)$, and the total structure factor, $S(Q)$. The starting configurations of a system containing 2000 As and 3000 Se atoms with a number density of 35.74 nm$^{-3}$ were generated using hard-sphere Monte Carlo simulations, i.e., random configuration excluding short bonds. In order to avoid unphysical atomic configurations, three constraints were applied. Firstly, the cut-off values were determined to be 0.235, 0.230, and 0.225 nm for the As-As, As-Se, and Se-Se atomic pairs, respectively, to avoid physically unreasonable spikes in the partial pair-distribution functions, $g_{ij}(r)$, in the low $r$ range. Secondly, weak bond angle constraints were applied for each type to be about 100$^\circ$. Finally, a constraint of a wrong Se-Se bond ratio was applied around the Se atoms. A ratio of 16.0 \% was chosen for this constraint, being determined from the result of the {\it ab initio} MD simulation.

The {\it ab initio} MD calculation was based on density functional theory with the generalized gradient approximation (GGA) \cite{Perdew} for the exchange-correlation energy. The projector augmented wave (PAW) potentials \cite{Blochl} were employed for the electron-ion interaction with valence electrons $4s^24p^54d^0$ and $4s^24p^64d^0$ for As and Se, respectively. The electronic wavefunctions and the electron density were expanded by plane wave basis sets  with cutoff energies of 11 and 55 Ryd, respectively. The $\Gamma$ point was only used to sample the Brillouin zone of the MD supercell. We used 128 atoms (48 As and 72 Se) in a cubic MD cell with periodic boundary conditions. The side length of the MD cell was 1.48 nm. The constant-temperature MD simulation \cite{Nose,Hoover} was performed at 300K for 30,000 steps in time steps of 2.9 fs. The quantities of interest were obtained by taking average over ten simulations starting from different initial configurations taken from simulations for the liquid state at 800 K and different cooling rates, where only small dfferencies were found in the reslts.

\begin{figure*}
\onefigure[width=140mm]{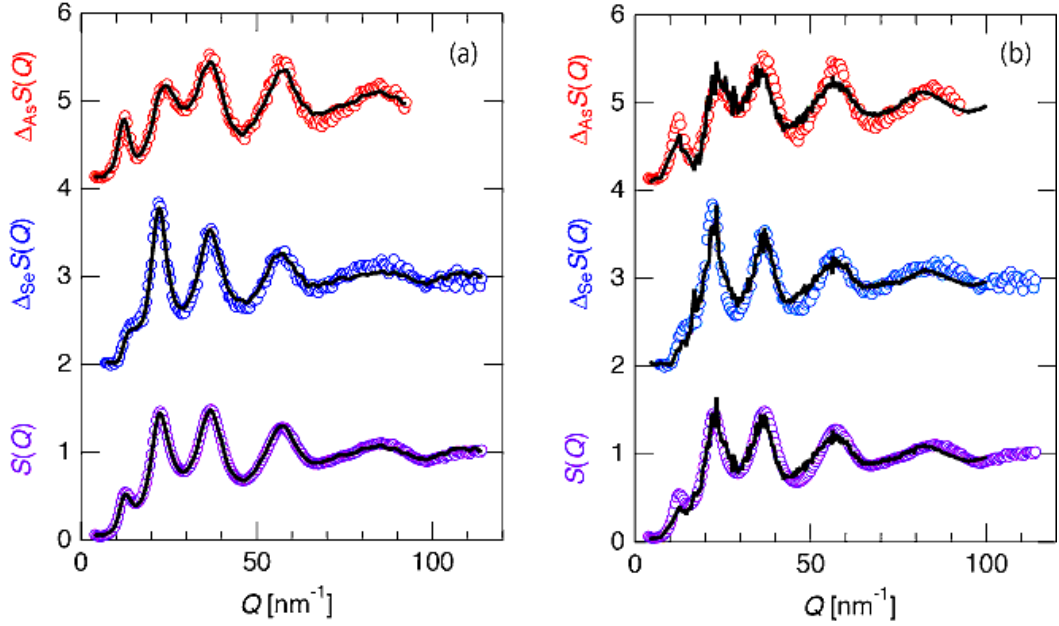}
\caption{(Color online.) From top to bottom: $\Delta_kS(Q)$ around the As and Se $K$ edges, and $S(Q)$. Circles indicate those obtained from the AXS experiments, and solid curved denote (a) fits by the RMC modeling and (b) results calculated from the {\it ab initio} MD simulation.}
\label{f1}
\end{figure*}

%\section{Results}
Circles in Fig. \ref{f1} show $\Delta_kS(Q)$ obtained from the present AXS measurements close to the As and Se $K$ edges, together with $S(Q)$. $\Delta_{\rm As}S(Q)$ has a distinct and sharp prepeak at about $Q=12$ nm$^{-1}$, and the $Q$ position almost coincides with that in $S(Q)$, although the peak height of the latter is very small. $\Delta_{\rm Se}S(Q)$ shows a shoulder at a higher $Q$ value of about 14.5 nm$^{-1}$. Solid curves in Fig. \ref{f1}(a) show the best fits of the RMC atomic modeling, which mostly coincide very well with all of the present experimental data. 

Solid curves in Fig. \ref{f1}(b) indicate results of the {\it ab initio} MD simulation, which were calculated from theoretically obtained partial structure factors, $S_{ij}(Q)$, by taking atomic form factors for x-ray scattering into account. The circles denote again the experimental data. Overall features of the theoretical spectra are in good agreement with the experimental results although the prepeaks and the shoulder in the low $Q$ region are slightly smaller. 

\begin{figure}
\onefigure[width=70mm]{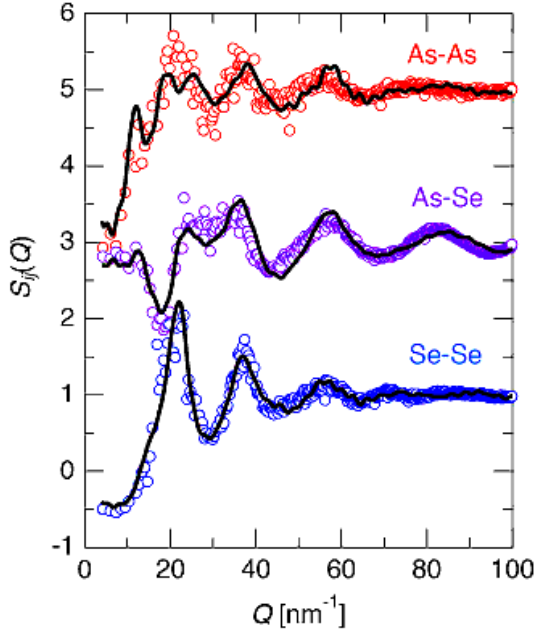}
\caption{(Color online.) The $S_{ij}(Q)$ functions obtained from the RMC modeling (solid curves) and the {\it ab initio} MD simulation (circles). 
}
\label{f2}
\end{figure}

Figure \ref{f2} shows $S_{ij}(Q)$s obtained from the AXS+RMC (solid curves) and the {\it ab initio} MD simulation (circles). The $S_{ij}(Q)$ results obtained from AXS+RMC and {\it ab initio} MD coincide quite well although the theoretical data are rather scattered. In $S_{\rm AsAs}(Q)$, a large and sharp prepeak locates at about 12 nm$^{-1}$, indicating that the intermediate-range order in this glass is mainly due to the As--As correlation as was pointed out  in Ref. \cite{HosokawaAs2Se3}. A small prepeak is also seen in $S_{\rm AsSe}(Q)$ at the same $Q$ position, and a steep hump is observed at nearly the first peak position in $S(Q)$. In $S_{\rm SeSe}(Q)$, there is no indication for a prepeak. Instead, a shoulder is observed at about 14.5 nm$^{-1}$. These features are very similar to those found in GeSe$_2$ glass \cite{HosokawaGeSe2}.

\begin{figure}
\onefigure[width=70mm]{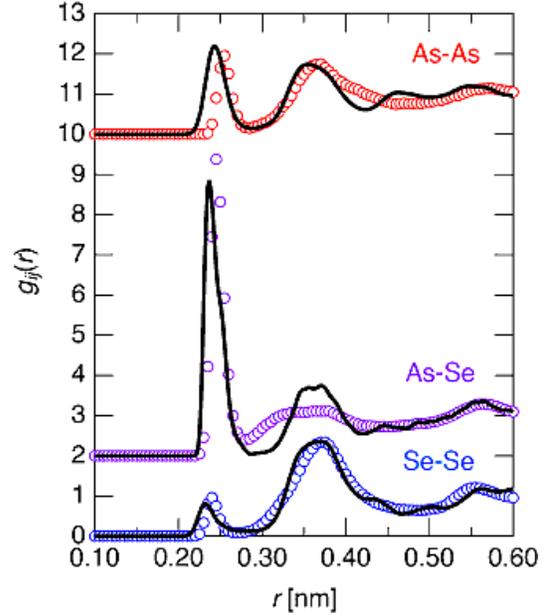}
\caption{(Color online.) The $g_{ij}(r)$ functions obtained from the RMC modeling (solid curves) and the {\it ab initio} MD simulation (circles).
}
\label{f3}
\end{figure}

Figure \ref{f3} shows $g_{ij}(r)$s obtained from the AXS+RMC (solid curves) and the {\it ab initio} MD simulation (circles). The first peak in the total pair-distribution function (not shown) mainly results from the As-Se heteropolar bonds centered at 0.237(1) nm (AXS+RMC) and 0.246(2) nm ({\it ab initio} MD), which are in good agreement with previous total diffraction works \cite{Renninger}.

The As-As and Se-Se homopolar bonds, the so-called wrong bonds do exist at nearly the same $r$ position as the As-Se heteropolar bonds. The existence of such homopoler bonds in As$_2$Se$_3$ glass was theoretically predicted by Li and Drabold \cite{Li} and more recently by Bauchy and Micoulaut \cite{Bauchy}. To our knowledge, however, this is the first experimental proof of the existence of such wrong bonds in the As$_2$Se$_3$ bulk glass. The peak heights of the first peaks in both the homo- and heteropolar bonds coincide well between experiment and theory, while the peak positions systematically shift by about 0.01 nm. 

%\section{Discussion}
The partial coordination numbers, $N_{ij}$, were obtained from the $g_{ij}(r)$ functions. Here, $i$ and $j$ indicate the central and neighboring elements, respectively. They were estimated by the integrals of $4\pi r^2g_{ij}(r)$ up to the minima of the theoretical $g_{ij}(r)$ functions, i.e., 0.290, 0.280, and 0.260 nm for the As-As, As-Se, and Se-Se correlations, respectively. The results are given in Table \ref{Nij}. $N_{ij}$ show good agreement between experiment and theory, except for the As-As correlation. 

\begin{table}
\caption{Partial and total coordination numbers obtained from experiment (AXS+RMC) and theory ({\it ab initio} MD simulation).}
\label{Nij}
\begin{center}
\begin{tabular}{lllll}
\hline
&& Neighbor  \\
&Central & As & Se & Total  \\
\hline
experiment & As & 0.73(1) & 2.53(1) & 3.26(2) \\
&Se & 1.69(1) & 0.32(1) & 2.01(2) \\
\hline
theory & As & 0.53(2) & 2.54(1) & 3.07(3)  \\
&Se & 1.69(1) & 0.32(1) & 2.01(2) \\
\hline
\end{tabular}
\end{center}
\end{table}

The total coordination numbers, $N_i$, are also given in Table \ref{Nij}. The $N_{\rm Se}$ values are two from both the experimental and theoretical results  within the errors, indicating that the `8-$N$ bonding rule' is strictly valid around the Se atoms. On the other hand, the $N_{\rm As}$ values exceed this octet rule value of three in experiment and theory although with a different extent: the AXS+RMC result indicates about 26\% of fourfold coordinated As atoms on average, while the {\it ab initio} MD simulation shows about 7\%. 

To examine the constraint effect of the ratio of homopolar Se-Se bonds around the Se atoms (16\%) in the RMC calculation, we carried out several RMC calculations with ratios from 0 to 50\%. As a result, the same $N_i$ values were obtained within the error bars given in Table \ref{Nij}. Another RMC calculation was performed with a different constraint of no homopolar As-As wrong bonds, and the same $N_i$ values were obtained producing As(Se$_{1/2}$)$_4$ tetrahedra, whereas the RMC fit functions largely deviated from the experimental structure factors. We also examined the the initial configuration dependence by using fully random configurations or crystal structures. Consequently, the same results were obtained in $S_{ij}(Q)$, $g_{ij}(r)$, and $N_{ij}$. 

Another doubt arose about small discrepancies between $\Delta_kS(Q)$s and the RMC fits in the high $Q$ region shown in Fig. \ref{f2}(a). We think that these discrepancies originate from the experimental errors. As is well-known, atomic form factors in x-ray diffraction (XD) signals rapidly decrease with increasing $Q$, and the experimental errors becomes large at the high $Q$ values, in particular, in $\Delta_kS(Q)$s, because they are contrasts of the XD data of several \% with changing incident x-ray energies. Also, the RMC is reflected by atomic configurations in real pace, but the experimental data are not. Moreover, the larger oscillations in $\Delta_kS(Q)$s do not result in the decrease of $N_{\rm As}$. Moreover, the RMC and {\it ab initio} MD results are in better agreement with each other in the high $Q$ range. Thus, the over-coordinated feature around As should not be broken down by such experimental errors.

A previous {\it ab initio} MD simulation on As$_2$Se$_3$ glass was performed by Li and Drabold \cite{Li}. They found the existence of the wrong bonds, $N_{\rm AsAs}=0.67$ and $N_{\rm SeSe}=0.43$. Compared with the present results, the former is in good agreement, and the latter is slightly larger. A discrepancy is seen in $N_i$ that As and Se atoms have 3.01 and 1.99 neighbors on average in the first shell, respectively, indicating the validity of `8-$N$ bonding rule' within the errors. They did not describe the explicit definition of $N_{ij}$, and only wrote that all atoms in the first shell were distributed in the range 0.234--0.263 nm. If they calculated the $N_{ij}$ values up to 0.263 nm, the obtained values $N_{ij}$ may be underestimated, in particular, for $N_{\rm AsAs}$, since a large tail of the first peak is seen in $g_{\rm AsAs}(r)$ at $r=0.263$ nm in Fig. 3 of Ref. \cite{Li}.

Recently, another {\it ab initio} MD simulations on As$_2$Se$_3$ glass and other As-Se glasses were performed by Bauchy and Micoulaut \cite{Bauchy}. They confirmed the existence of wrong bonds in As$_2$Se$_3$ glass, $N_{\rm AsAs}=0.64$ and $N_{\rm SeSe}=0.42$, in good agreement with theory including the present one and the present experiment. It should be noted that 9.2 \% of fourfold coordinated As atoms are found in the As$_2$Se$_3$ glass, resulting in a number of constraints per atom to be 3.05, well exceeding three, the degree of freedom in a three-dimensional system.

\begin{figure}
\onefigure[width=70mm]{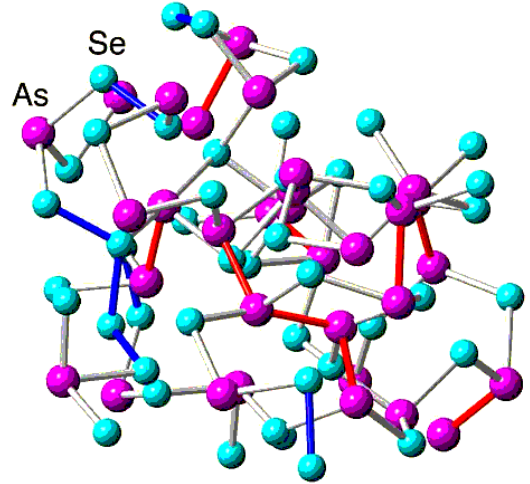}
\caption{(Color online.) A snapshot of 3D atomic configuration obtained from the {\it ab initio} MD simulations.
}
\label{f4}
\end{figure}

Next, we discuss local atomic configurations or structural motifs in the As$_2$Se$_3$ glass. Figure \ref{f4} shows a snapshot of the 3D atomic configuration obtained from the {\it ab initio} MD simulation. Large purple and small green balls indicate the As and Se atoms, respectively. The As-As  and Se-Se wrong bonds are given as dark red and blue bars, respectively. Both fourfold- and  threefold coordinated As atoms are observed. Of special interest is that around the fourfold coordinated As atoms, the atomic configuration is not always Se$_3$--As--As--Se$_3$ ethan-like, but other As neighbors are frequently seen, i.e., As-As bond chains are formed. 

\begin{figure}
\begin{center}
\onefigure[width=70mm]{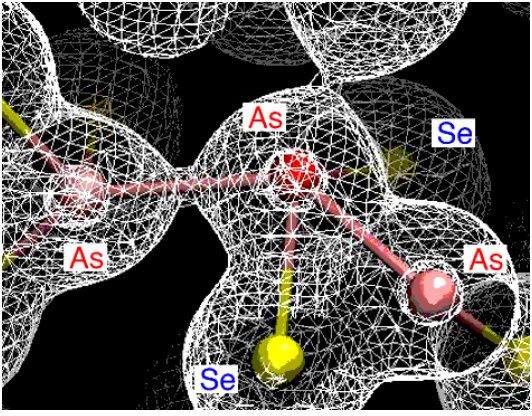}
\caption{(Color online.) An example of the spatial distribution of the electron charge density around a fourfold As atom in a As-As wrong bond chain.
}
\label{f4a}
\end{center}
\end{figure}

Figure \ref{f4a} shows an example of the spatial distribution of the electron charge density around a fourfold As atom in a As-As wrong bond chain. The central fourfold As atom has two Se and two As neighbors. Since the bond distance with the As atom of the left-hand side is longer than the average, the bonding electrons are very few in number. Nevertheless, the existence of covalent electrons is confirmed for every bonds. From the further discussion using bond-overlap population (BOP) analysis \cite{Shimojo}, two As-Se bonds around the fourfold As atoms mostly form strong bonds with two covalent electrons, and are rarely broken during the MD simulation time. On the other hand, two As-As bonds have a tendency to share the remaining one covalent electron with each other, and the bonds are short-living with the lifetime of some ps. Note that the strengths of two As-As bonds are different depending on the environment surrounding the two bonds. Due to the change of the bond length caused by the lattice vibration, repeated switchings of the bonding electron are observed between the right and left As-As bonds in some hundred fs steps.

\begin{figure}
\onefigure[width=70mm]{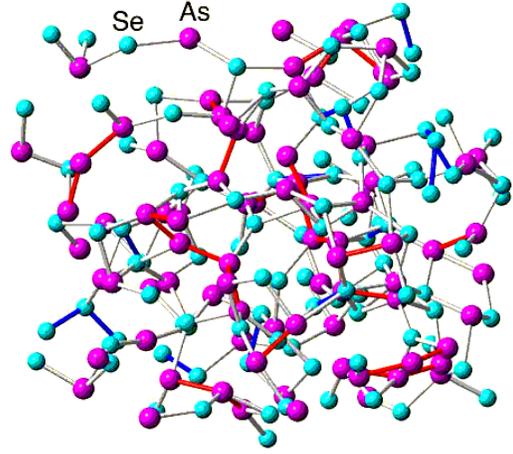}
\caption{(Color online.) An example of 3D atomic configuration obtained from the AXS+RMC results.
}
\label{f5}
\end{figure}

Figure \ref{f5} shows an example of 3D atomic configuration obtained from the AXS+RMC. The marks are the same as in Fig. \ref{f4}. As clearly seen in the figure, this experimental result also shows the existence of the As-As wrong bond chains, but longer than in the theoretical result, i.e., a five- or six-atoms chain was observed in the atomic configuration. From these findings, it is concluded that the fourfold coordinated As atoms likely form As-As wrong bond chains rather than ethan-like configurations. 

Then, two new questions arose. The first one is why the significant changes of the local environment around the As atoms do not affect the electronic states, in particular, isolated electron spins detectable with ESR experiment \cite{Bishop} or electron distributions observable by NMR measurement \cite{Hari}. We believe that this may be solved by the above results of BOP analysis. The number of electrons belonging originally to the fourfold As atoms remains unchanged to be three. Each electron forms a pair with an electron from the neighboring As or As atom as a usual As(Se$_{1/2}$)$_3$ pyramid. Thus, no difference would be seen by the ESR and NMR measurements. The difference in the present fourfold As tetrahedron is that one electron is shared with two As-As bonds and a fast flickering of the electron positions may occur between two bonds. Also, this mechanism is the reason why the As-As wrong bond chains are preferable rather than ethan-like configurations with single As-As bond.

The second question is why there is a large discrepancy between the $N_{\rm As}$ values in the experiment (26\%) and the theory (7\%), even though they exceed the `8-$N$ bonding rule' value of three. The wrong bond chains may explain it. Since the MD simulation box size is about five atoms, it acts as a constraint avoiding the longer As-As wrong bond chains even in the periodic boundary condition. Atom number dependence was examined in the MD simulation using smaller numbers of atoms of 64 with a smaller box size. Unfortunately, we saw only a small increasing trend in the average coordination number around the As atoms with increasing number of atoms in the calculation within the error bar. The recent {\it ab initio} MD simulation by Bauchy and Micoulaut, however, showed more than 9\% of fourfold coordinated As atoms using a larger number of atoms of 200, which suggests that the above speculation would not be ridiculous. By taking the As-As chain length obtained from the AXS experiment with the RMC model into account, however, more than 2000 atoms would be necessary to reproduce the 26\% of fourfold coordinated As atoms as obtained from the experiment. However, it is unrealistic at present. 

Compared to the prediction \cite{Georgiev} based on the rigidity percolation theory, the portion of the fourfold coordinated As atoms, 28.6\%, is in good agreement with the present AXS+RMC result of about 26\%. However, they predicted the Se=As(Se$_{1/2}$)$_3$ quasi-tetrahedral units for the fourfold coordinated As atoms, while our experimental and theoretical data show the As-As wrong bonds in the network glass. 

In summary, this is the first experimental proof for the breakdown of the `8-$N$ bonding rule' proposed by Mott \cite{Mott1969,MottBook}. It was only seen around the As atoms, while the Se atoms strictly keep this rule. This experimental result could be confirmed by an {\it ab initio} MD simulation, where the dynamic properties such as the lifetime of the fourfold coordinated As atoms, i.e., a flickering motion of the As-As bonds around the fourfold coordinated As atoms, were discussed together with the corresponding electronic natures, which do not contradict the results of ESR \cite{Bishop} and NMR \cite{Hari} measurements. The general picture of the averagely over-coordinated As atoms is valid in the As$_2$Se$_3$ glass, and no crucial failures were found against this conclusion at present as discussed in this paper in detail. 

However, readers of this paper may still wonder about the validities of the experiment (AXS+RMC) or the theory ({\it ab initio} MD) individually. For example, investigations of structures of non-crystalline materials have large ambiguities in the results although the accuracies of scattering techniques have recently been improved drastically by using synchrotron radiation or new intense neutron sources. RMC modeling is only an example of atomic configurations of non-crystalline materials, even if the model fits properly to experimental data. 

The box size for {\it ab initio} MD simulations is always a subject of the criticisms, which cannot be overcome using a limited computer power at present. Moreover, too short calculation time of some ps does not correspond to the real relaxation time of glasses from melts of ms or longer. Atomic configurations  obtained from {\it ab initio} MD simulations are possible to be still those in the melts. As pointed out by Bauchy and Micoulaut \cite{Bauchy}, therefore, one has to be careful with the proposed statistics of the corresponding fourfold coordinated As fraction could be different and affected by the computer timescale.  

However, one should composedly judge how these limitations influence our final conclusion of the {\it average} coordination number around the As exceeding the `8-$N$ bonding rule' of three. On the experimental side, the intermediate-range atomic configurations are surely depend on the wrong bond ratio around the Se atoms, while the $N_{\rm As}$ value does not vary with such constraints in the RMC modeling. On the theoretical side, the density of the system already reaches the actual relaxed value under ambient conditions during the short time of some ps. The concentration of the wrong bonds may change by a longer and slower relaxations from melts to glasses. As regards the average coordination numbers, however, it is difficult at present to decide whether the values increase, decrease, or remain unchanged.

Here, we emphasize that a good agreement of the over-coordinate feature around the As atoms was obtained from two individual methods of experiment and theory, which is, we think, beyond an accidental coincidence, although both the methods have their own limitations. We could not find any failures against our conclusion of a breakdown of the `8-$N$ bonding rule' around the As atoms in the As$_2$Se$_3$ glass at present.

\acknowledgments
The AXS experiments were performed at the beamline BM02 of the ESRF (Proposal Nos. HS1860 and HS2184).

\end{document}